\documentclass[twocolumn,aps,raggedbottom,nobalancelastpage,amssymb,floatfix,superscriptaddress,nofootinbib]{revtex4-1}

\usepackage{graphicx}
\usepackage{amsmath}
\usepackage{amssymb}
\usepackage{bm}
\usepackage{pgf}
\usepackage{color}
\usepackage{amsfonts}
\usepackage{hyperref}
\hypersetup{colorlinks = true, linkcolor=magenta,citecolor=blue, urlcolor=magenta, bookmarksnumbered =  true}
\usepackage{siunitx}

\newcommand{\ssm}{\rm\scriptscriptstyle}

\begin{document}
	
\title{Intrinsically polar elastic metamaterials}
\author{Osama R. Bilal}
\affiliation{Department of Mechanical and Process Engineering, ETH Zurich, 8092 Z\"urich, Switzerland}
\affiliation{Institute for Theoretical Physics, ETH Zurich, 8093 Z\"urich, Switzerland}
\affiliation{Division of Engineering and Applied Science, California Institute of Technology, Pasadena, CA 91125, USA}
\author{Roman S\"usstrunk}
\affiliation{Institute for Theoretical Physics, ETH Zurich, 8093 Z\"urich, Switzerland}
\author{Chiara Daraio}
\affiliation{Division of Engineering and Applied Science, California Institute of Technology, Pasadena, CA 91125, USA}
\author{Sebastian D. Huber}
\affiliation{Institute for Theoretical Physics, ETH Zurich, 8093 Z\"urich, Switzerland}

\begin{abstract}
 
The ability to design and fabricate materials with tailored mechanical properties, combined with immunity to damage, is a frontier of materials engineering. For example, materials which are characterized by elastic properties that depend on the position inside a medium are required in applications where structural stability has to be combined with a soft and compliant surface, like in impact protection and cushioning. A gradient in the elastic properties can be built from a single material, varying gradually the bulk porosity of the material or its geometrical structure. However, if such a gradient is built into the material at production, damage or wearing over time might expose unwanted elastic properties. Here, we implement a design principle for a spatially inhomogeneous material based on topological band-theory for mechanical systems. The resulting inhomogeneity is stable against wearing and even cutting the material in half. We show how, by creating a periodic elastic material with topological properties, one can create an intrinsically polar behavior, where a face with a given surface normal is stiff while its opposing face is soft.\footnote{This is the pre-peer reviewed version of the following article: Adv. Mater. 1700540 (2017), which has been published in final form at \href{https://dx.doi.org/10.1002/adma.201700540}{URL}. This article may be used for non-commercial purposes in accordance with Wiley Terms and Conditions for Self-Archiving.} 
\end{abstract}

%TC:endignore 

\maketitle

\begin{figure}[!b]
  \begin{center}
    \includegraphics{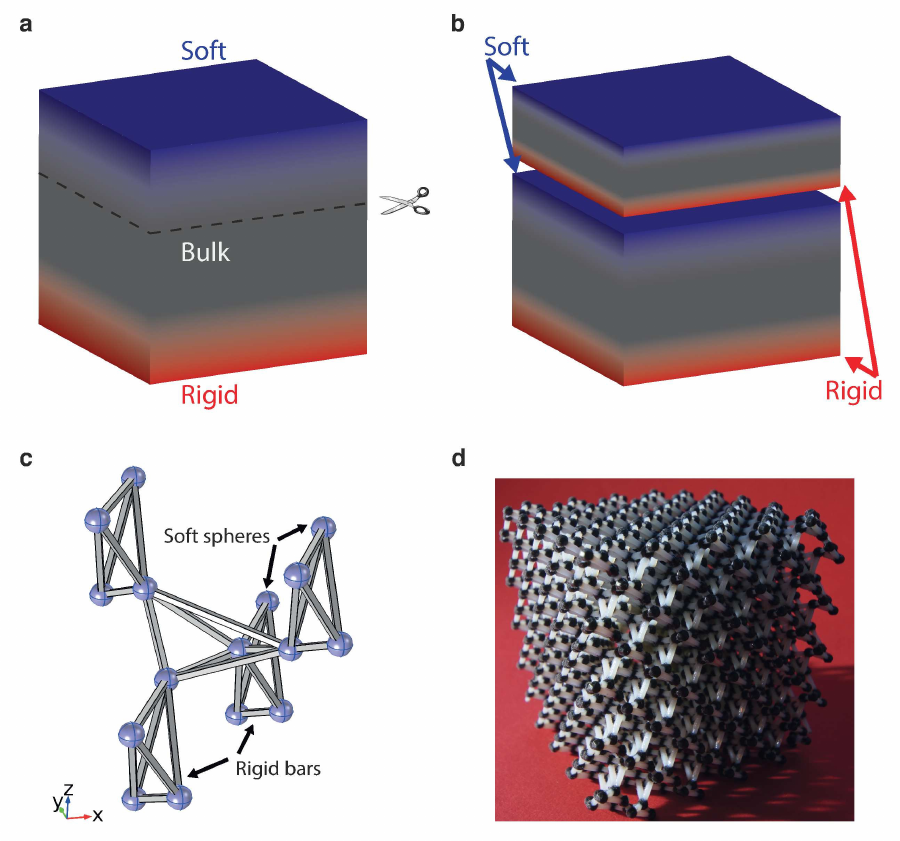}
  \end{center}
      \caption{{\bf Intrinsically polar metamaterial.} (a) Schematic representation of an elastically polarized mechanical metamaterial with varying stiffness from top to bottom surface. (b) Due to the intrinsic programmability of its elastic properties, whenever a cut is performed through the material, the polarity of the surfaces are protected by means of topology, retaining the variation of stiffness from top to bottom. (c) Schematic of the basic building block of the employed Pyrochlore lattice. (d) a $5\times 5\times 5$ unit cells of the realized metamaterial by means of additive manufacturing.  }
  \label{fig:schematic}
\end{figure}

The characteristic properties of periodic materials are captured in their phononic spectrum, i.e., their dispersion relation, which contains information ranging from the material's quasi-static elastic response (at very low frequencies) to its thermal conductivity (at higher frequencies). In topological (polar) materials, the phonon spectrum, and with it the linked material properties, are different on two opposing surfaces of the material. Uniquely to topological materials, the polar behavior is stable and remains preserved when these materials are cut or fractured, as shown in Fig.~\ref{fig:schematic}a,~b:  No matter how much of our material is lost to a fracture, or wearing during use, the asymmetry in the mechanical response remains the same \cite{Kane13, Huber16, Stenull16, Susstrunk16}. Note, that these topological features are linked to surface modes whose penetration depth introduces a new length-scale and hence the stability with respect to wearing and fracturing is not bound to the microscopic unit cell size.  This similarity to the polarity of a dielectric motivates the term of an {\em intrinsically polar elastic metamaterial}. 

As it turns out, the simplest lattice that lends itself to the design of such a polar metamaterial has another topologically protected feature in its phononic spectrum: Lines of zero-frequency excitation that span throughout the whole Brillouin zone and which cannot be gapped out \cite{Po16}. Such nodal, or Weyl lines \cite{Stenull16, Yang16, Po16, Rocklin16, Bernevig15} are of considerable recent interest in electronic systems \cite{Bernevig15} as they mediate non-local magneto-transport. In our mechanical setup, we find clear experimental evidence of these Weyl lines and we further capitalize on their presence to fine tune the elastic response. 

\begin{figure*}[t]
    \begin{center}
      \includegraphics{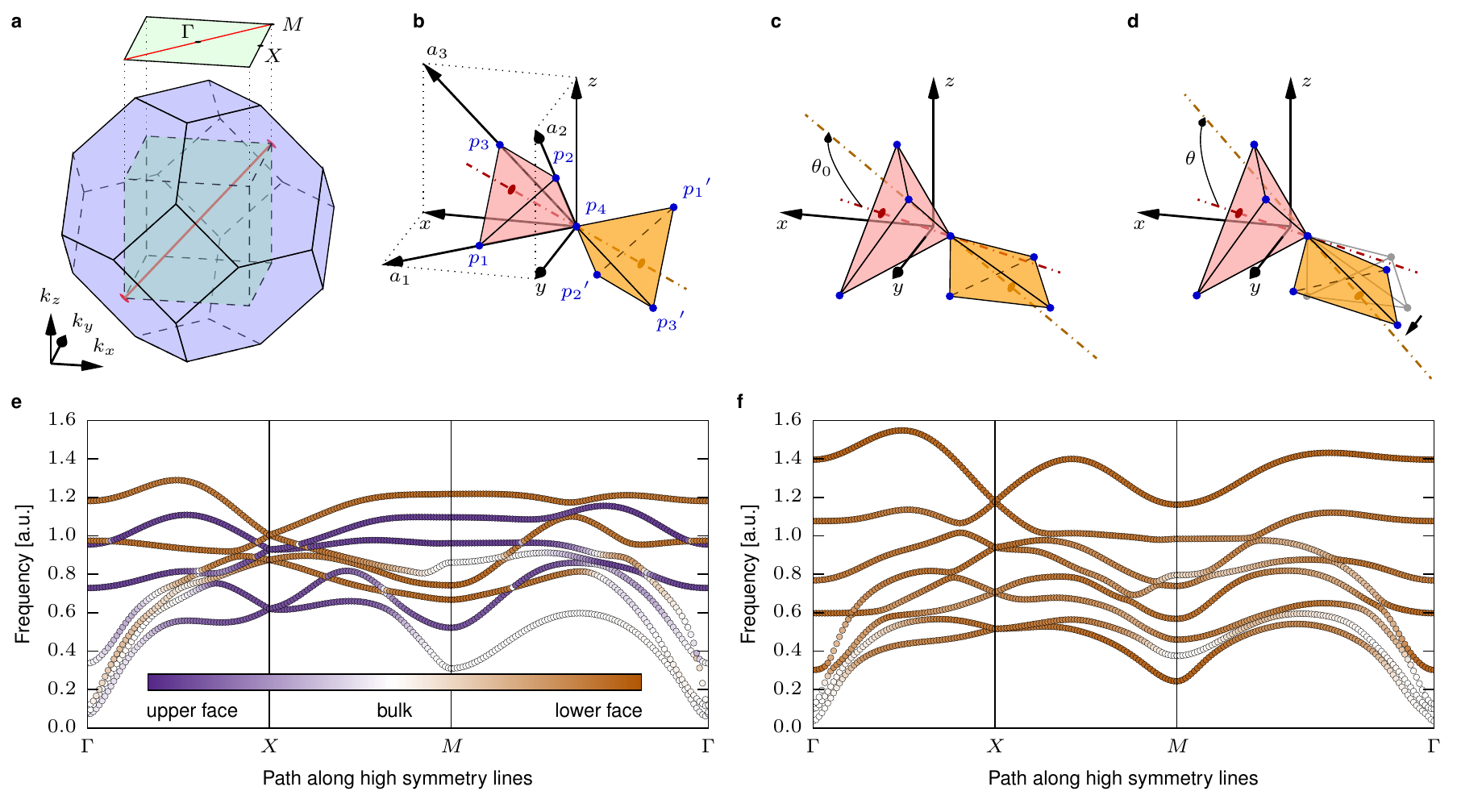}
    \end{center}
    \caption{{\bf Theory.} (a) Bulk (blue) and surface (green) Brillouin zone for a surface plane with a surface normal $\hat {\boldsymbol{z}}$. The red line shows the doubly degenerate nodal Weyl lines in the bulk spectrum. (b) Geometry of a regular corner-sharing Pyrochlore lattice. (c) The distorted Pyrochlore lattice. $\theta_0$ indicates the equilibrium angle between two neighboring tetrahedra. (d) In a zero mode of the perfect frame no bars are contracted or stretched. However, the angle between rigid tetrahedra can change from $\theta_0 \to \theta$. We model non-perfect hinges by a linear force   $\theta_0-\theta$. (e) Effective phonon spectrum in the zero-mode sector of the perfect frame as a function of high-symmetry lines in the surface Brillouin zone for a finite system in the $y$-direction. The color of the dots indicates where the mode lives (bottom or top surface or in the bulk). One can observe that the effective low-frequency modes are distributed equally over the two opposing surfaces. (f) The same plot for open faces in the $z$-direction. One can clearly see that no zero modes live on the upper face. 
    }
        \label{fig:theory}
\end{figure*}

We design our polar elastic materials as a truss-like, periodic lattice, constituted by a system of rods connected by hinges (a frame). In the classical description of frames, one balances the number of degrees of freedom ($N_f$) of the hinges with the number of constraints ($N_c$) imposed by the rods. Their difference yields the number of zero modes ($N_0$), where parts of the system can move freely without a restoring force. More precisely, the counting due to Maxwell \cite{Maxwell64} and Calladine \cite{Calladine78}
\begin{equation}
  N_0 - N_s = N_f - N_c
\end{equation}
accounts also for the number of states of self stress ($N_s$). A state of self stress corresponds to a combination of stresses on the rods that do not exert net forces on the hinges. In isostatic frames, where $N_f=N_c$, one might still find pairs of zero modes and states of self stress due to ``misplaced'' rods. 

In the theory of Kane and Lubensky \cite{Kane13,Lubensky15}, isostatic frames are described by a polarization vector $\boldsymbol{R}_{\ssm T}$ akin the electric polarization of dielectrics. The polarization $\boldsymbol{R}_{\ssm T}$ can be expressed as a topological invariant of the bulk of a periodic frame and is therefore stable against local deformations of the frame. Moreover, if a periodic isostatic lattice is cut to obtain a finite sample, one necessarily cuts bonds and zero modes will appear on the surface. The polarization $\boldsymbol{R}_{\ssm T}$ indicates how these zero modes are distributed over the different surfaces. This renders isostatic frames an optimal starting point for a polar metamaterial. However, we set out to design an asymmetric {\em elastic response}, whereas a zero-mode does not lead to any elasticity as there are no restoring forces. 

The idealized description of a frame as made out of perfect hinges will always be an approximation. The connections between the rods typically induce further constraints due to friction or elastic forces that favor a certain angle between the connected rods. We show how one can capitalize on this additional forces to obtain an elastic polar response. In our material, we use a stiff polymer to fabricate the rods, and a soft, elastic rubber for the hinges, and use additive manufacturing for the final fabrication of the composite material. We use this large separation of scales between rods and hinges, to derive an effective elastic description of our material, by projecting the angle-forces to the space of zero modes of the idealized frame. 

Isostaticity of a periodic frame in three dimensions requires each hinge to be connected to six bars. The simplest regular lattice with this coordination is the pyrochlore lattice, which is built from corner sharing tetrahedrons.  In order to leverage a polar response and to use the topological theory by Kane and Lubensky \cite{Kane13}, we distort the lattice as indicated in Fig.~\ref{fig:schematic}c (see App.~\ref{sec:structure}). The distorted pyrochlore lattice has the additional feature of two nodal Weyl lines of zero frequency excitations in its bulk spectrum, see Fig.~\ref{fig:theory}a. In principle, these Weyl lines render the topological polarization ${\boldsymbol R}_{\ssm T}$ ill-defined and in turn give rise to a number of zero modes on the surface that depends on the surface momentum \cite{Stenull16}. However, by designing the distortion in a way that some of the flat planes of the original pyrochlore lattice prevail, one can force the two Weyl lines to lie on top of each other along the $(1,1,1)$-direction, see Fig.~\ref{fig:theory}. This eliminates their effect on $\boldsymbol{R}_{\ssm T}$ and we can still achieve a maximal polarization between two opposing faces. In particular, for our choice of distortion, $\boldsymbol{R}_{\ssm T}=s(-1,0,2)$ with $s$ the overall scale. This results in a polarization where a face with a surface normal in the positive (negative) $z$-direction hosts four (no) zero modes. For faces normal to the $y$-direction there is a balance of two and two zero modes, cf. App.~\ref{sec:count} for details. 

The design principle of Maxwell frames mandates the presence of a rotating degree of freedom (i.e., a perfect hinge) at each intersection point between frame elements (bars). To design a material following this principle, one needs to carry out a nontrivial assembly process on the macro-scale. Moreover, the miniaturization of such procedure is tedious and impractical. In order to avoid the assembly process, while retaining a hinge-like behavior, we follow a different fabrication process. The realization of the metamaterial starts by separating the lattice into two distinct yet interconnected object classes, the bars and the joints. The bars are made out of beams of varying length and constant square cross section of width $1.125\,\text{mm}$. The joints are replaced by spheres of radius $1.5\,\text{mm}$. Both beams and spheres are fabricated using additive manufacturing technology (Polyjet 3D printing), that enables the realization of a single structure with multiple materials simultaneously. We harness this technology to obtain a hinge-like performance as the theory requires, by printing each of the different objects (spheres and beams) out of different materials. In order to ensure flexibility at the joint site, the spheres are printed with a much softer material, TangoBlack, (with density, $\rho= 1.15\, \text{g}/\text{cm}^3$ and Young's modulus $E_{\ssm TB} = 1.8\,\text{MPa}$) than the beams, VeroWhite, ($\rho = 1.17\, \text{g}/\text{cm}^3$, Young's modulus $E_{\ssm VW}= 2\, \text{GPa}$) with $10^3$ times lower stiffness and similar densities, which we will treat as being equal in the following. While the soft TangoBlack enables a description in terms of perfect hinges, its non-vanishing $E_{\ssm TB}$ gives rise to restoring forces on the angles between the beams. We incorporate these forces in our generalization of the theory of isostatic lattice.

To find the effective elastic theory we first find the zero frequency excitations for the idealized case of perfect hinges. The equations of motion read
\begin{equation}
    \ddot{\boldsymbol{x}} = \mathcal{D}_{\ssm iso} \boldsymbol{x},
\end{equation}
where the vector $\boldsymbol{x}$ contains the degrees of freedom of the hinges and $\mathcal{D}_{\ssm iso}$ encodes the effect of the bars and is proportional to $E_{\ssm VW}/\rho$. The eigenvectors of $\mathcal{D}_{\ssm iso} $ separate into two classes
\begin{equation}
    M_0 = \{v_1,v_2,\dots,v_z\} \; \mbox{and} \; M_{\perp} = \{v_{z+1},v_{z+2},\dots\;\},
\end{equation}
of $z$ zero modes in $M_0$ and its complement $M_{\perp}$. Any distortion of the lattice involving modes from $M_{\perp}$ stretches or compresses the bars. For the derivation of a low-frequency sector, we now proceed by projecting the angle-restoring forces $\mathcal{D}_{\ssm angle}$ onto the zero mode subspace
\begin{equation}
    \mathcal{D}_{\ssm eff} = M_{0}^{\rm T} \mathcal{D}_{\ssm angle} M_{0}.
\end{equation}
Note that any small deformation of the lattice is now giving rise to an elastic response encoded by $\mathcal{D}_{\ssm eff}$, which is proportional to $E_{\ssm TB}/\rho \ll E_{\ssm VW}/\rho$. The resulting low-energy phonon spectrum is shown in Fig.~\ref{fig:theory}. In panel e, we show the effective low-frequency spectrum for a system periodic in $x$ and $z$ direction and of finite extent in the $y$ direction. From the color code we read that on top of the Weyl bulk modes (lifted to finite frequencies owing to the angle-forces) we have as many low-energy states on the bottom and the top face. In panel f the same situation is shown where $x$ and $y$ direction are periodic. Clearly, all surface zero modes are concentrated on the lower face. Note that the penetration depth of the surface zero-modes decreases with the amount of distortion and can be much larger than the size of a unit cell \cite{Stenull16}.

\begin{figure}[ht]
	\begin{center}
		\includegraphics{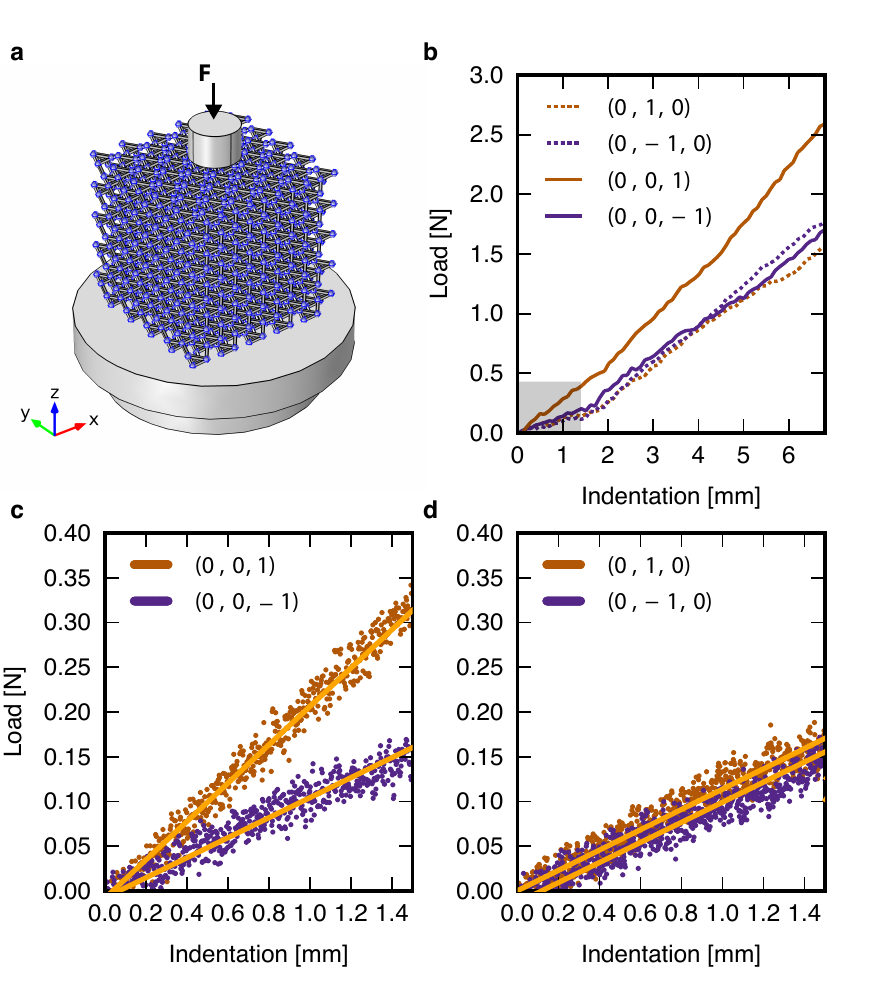}
	\end{center}
	\caption{\textbf{Measured elastic polar response.} (a) Schematic representation of the experimental setup for an asymmetric load indentation. (b) Measured response of a cube made out of $5\times5\times5$ unit cells to large indentation on faces along both $y$ and $z$ axes. The gray region highlights the range of linear (small) response. Measured linear response of the six faces of the lattice cube in (c) $z$ direction and (d) $y$ direction.}
	\label{fig:point-load}
\end{figure}

In a cube made out of an isotropic material, such as most metals, the stiffness measured on its different faces are identical. If the material is anisotropic, such as wood or fibers, one should expect directional dependence in stiffness (i.e., much stronger along the fiber than across it). In either case, isotropic or anisotropic, the stiffness along opposing faces is the same, in other words the material response along the same axis is symmetric. In order to test the asymmetric response of the proposed lattice material, we fabricate a cubic sample made of $5\times5\times5$ unit cells. Using a standard compression testing machine (Instron E3000), we indent the cube at the center of a particular face with a cylindrical probe ($16\,\text{mm}$ in diameter) moving for a fixed distance (Fig.~\ref{fig:point-load}a), while measuring the reaction force on the opposing face (Fig.~\ref{fig:point-load}b). We repeat the same process for all the faces of the cube. For the two faces along the $y$ axis, (0,$\pm$1,0), the material response to the same compression is almost identical (dashed lines in Fig.~\ref{fig:point-load}b). On the contrary, the two $z$ axis faces, (0,0, $\pm$1), where we expect the asymmetry, the response to the indentation is very different at the linear scale, and keeps on diverging with increasing indentation value, even in the nonlinear range (solid lines in Fig.~\ref{fig:point-load}b). Since the theory is only concerned with linear phonons, we focus on the small indentation region, the gray area in (Fig.~\ref{fig:point-load}b). To characterize the asymmetry of the lattice, we plot the response of each two opposing faces in the same panel (Fig.~\ref{fig:point-load}c,~d). In panel d, the slope of the response curve is identical along the $y$ axis, while in panel c, the two faces along the $z$ axis show a discrepancy in stiffness of $\sim$ 80\%. This provides an experimental evidence of the realization of an intrinsically polarized mechanical metamaterial with an asymmetric elastic response.

\begin{figure}[ht]
  \begin{center}
    \includegraphics{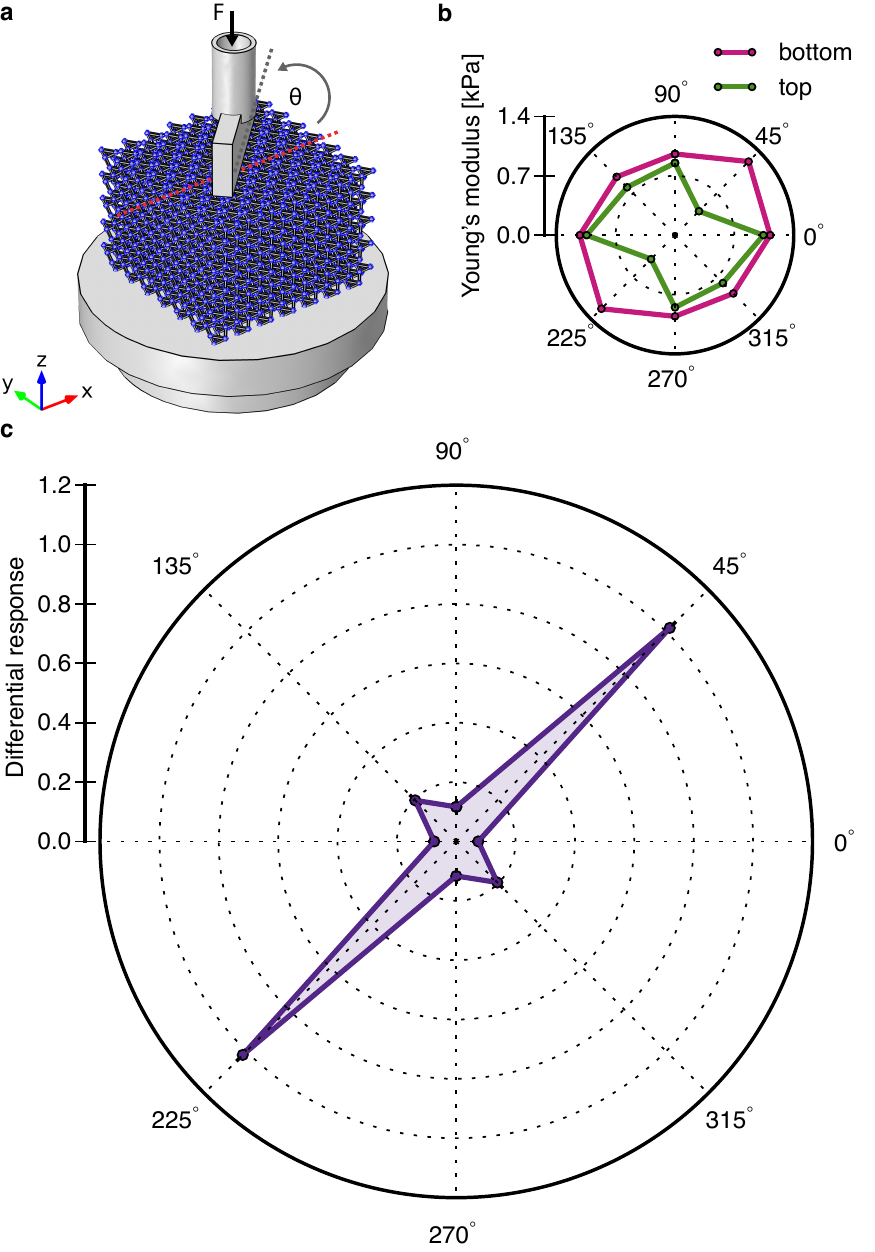}
  \end{center}
  \caption{\textbf{Observation of Weyl line.} (a) Schematic representation of the experimental setup for wedge-load indentation. (b) Obtained Young's moduli for bottom (pink) and top (green) surface indented at different angles spanning \ang{360}. (c) The percentage difference between the Young's modulus of the top and bottom surfaces.  Both top and bottom surfaces retain similar response to indentation at different angles except along \ang{45} and \ang{225}, where the indenter overlaps with the Weyl lines.}
  \label{fig:wedge-load}
\end{figure}

Let us turn to the effect of the Weyl nodal lines. We experimentally observe their effect by measuring the material response on the two opposing faces perpendicular to the $z$-axis by indenting with different planes parallel to this $z$-axis, cf. Fig.~\ref{fig:wedge-load}. When indenting with a plane, we pick up an elastic response from all modes with surface momenta $\boldsymbol{k}_{\perp}$ perpendicular to that plane, whereas along the plane direction one only gets weight from $\boldsymbol{k}_{\parallel}=0$. Therefore, when the plane of indentation is oriented {\em perpendicular} to the projection of the Weyl lines onto the surface Brillouin zone [the $(1,1)$-direction], we expect maximal participation of the bulk modes and hence a minimal difference between the two faces. Conversely, when indenting {\em parallel} to the Weyl lines, we should find a maximally different response.

In order to characterize this peculiar phenomenon of plane dependence, we print a lattice consisting of $7\times7\times5$ unit cells following the same fabrication process. We indent the printed lattice at the center with a rectangular wedge ($10\,\text{mm}$ $\times$ $80\,\text{mm}$) perpendicular to the $z$ axis, where the centers of both the wedge and the lattice top face coincide. After performing the compression test along the principle axis $x$ = 0, we rotate the wedge with an angle $\theta$ in a counterclockwise fashion in increments of \ang{45} and repeat the test until we reach a full circle in rotation (Fig.~4a). We repeat the same experiment for the bottom face of the sample along the $z$ axis as well. We post-process the compression-test data for the different angles, by calculating the slope of the indentation-load curve, to obtain the Young's modulus of the material and present it in a polar plot for both top and bottom faces in the $z$ direction (green and magenta lines in Fig.~\ref{fig:wedge-load}b). The Young's moduli on both surfaces at the same angle are similar, except for the plane along \ang{45} and \ang{225}. The normalized difference between the stiffness of the top and bottom surfaces at each angle is given in Fig.~\ref{fig:wedge-load}c. The measured differential stiffness in different planes across the lattice shows more than a factor 5 of variation on top and bottom surfaces along the same axis. This represents an experimental observation of nodal Weyl lines for phonons and a design methodology to intrinsically program exotic properties in materials.

We have realized a new class of lattice materials introducing a measurable intrinsic polarity in elasticity. The proposed lattice is material (metal, ceramic or polymer) and scale (micro or macro) independent, as it retains its property from structure instead of chemical compound. The elastic polarity can be coupled to dynamical, thermal, optical or electronic properties leading to the discovery of new materials with unprecedented properties. 

\begin{acknowledgements}

We acknowledge fruitful discussions with S.~Kroedel, T.~Lubensky, and V.~Vitelli and we thank T.~Jung for help with the additive manufacturing. ORB acknowledges support from the ETH Postdoctoral Fellowship FEL-26 15-2. SDH and RS are supported by the Swiss National Science Foundation.

\end{acknowledgements}

\appendix

\section{Lattice structure.}
\label{sec:structure}

For the lattice characterization we follow the description used in Ref.~\cite{Stenull16}. The unit cell of the distorted pyrochlore lattice is given by the position of its lattice sites $\boldsymbol{r}_i\!^s $ and the connecting bond centers $\boldsymbol{r}_i\!^b$. The lattice sites are obtained by distorting the lattice sites $\boldsymbol{p}_i $ of the ordinary pyrochlore lattice according to
\begin{equation}\label{eq:distortion}
	\begin{aligned}
		\boldsymbol{r}_1\!^s &= \boldsymbol{p}_1+x_1\sqrt{3}\hat{\boldsymbol{e}}_1-x_2\hat{\boldsymbol{a}}_3, \\
		\boldsymbol{r}_2\!^s &= \boldsymbol{p}_2+x_2\sqrt{3}\hat{\boldsymbol{e}}_2-x_3\hat{\boldsymbol{a}}_1, \\
		\boldsymbol{r}_3\!^s &= \boldsymbol{p}_3+x_3\sqrt{3}\hat{\boldsymbol{e}}_3-x_1\hat{\boldsymbol{a}}_2, \\
		\boldsymbol{r}_4\!^s &= \boldsymbol{p}_4-z \hat{\boldsymbol{n}},
	\end{aligned}
\end{equation}
where
\begin{equation*}
	\boldsymbol{p}_1 = \frac{s}{2} (1,1,0),\; \boldsymbol{p}_2 =\frac{s}{2} (0,1,1),\; \boldsymbol{p}_3 = \frac{s}{2} (1,0,1),\; \boldsymbol{p}_4 =\boldsymbol{0} \,,
\end{equation*}
with $s$ the overall scale, $\boldsymbol{a}_1 = \boldsymbol{p}_2-\boldsymbol{p}_1$,  $\boldsymbol{a}_2 = \boldsymbol{p}_3-\boldsymbol{p}_2$,  $\boldsymbol{a}_3 = \boldsymbol{p}_1-\boldsymbol{p}_3$, $\boldsymbol{e}_i = \boldsymbol{a}_i \times \hat{\boldsymbol{n}}$, $\boldsymbol{n}=(1,1,1)$, $\hat{\boldsymbol{v}} = \boldsymbol{v}/|\boldsymbol{v} |$ and $\boldsymbol{X}=(x_1,x_2,x_3,z)$ the parametrization. In our implementation we chose $X=0.15 s (-1,1,1,-1)$ and $s=7.5\,\text{mm}$. The lattice vectors defining the full lattice are $\boldsymbol{T}_i = 2 \boldsymbol{p}_i$ for $i=1,2,3$.

\section{Surface zero mode count.}
\label{sec:count}

The zero mode count $\nu$ per unit cell depends on the surface orientation. For a surface normal equal to a reciprocal lattice vector $\boldsymbol{q} $, it is given by \cite{Kane13}
\begin{equation}
	\nu = \frac{1}{2\pi} \boldsymbol{q}\cdot \bigl(\boldsymbol{R}_{\ssm T}+\boldsymbol{R}_{\ssm L}\bigr)\,,
\end{equation}
with the local dipole moment 
\begin{equation}
	\boldsymbol{R}_{\ssm L} = 3\sum_i \boldsymbol{r}_i\!^s - \sum_i \boldsymbol{r}_i\!^b
\end{equation}
and the topological polarization $\boldsymbol{R}_{\ssm T}$.

The latter is given by \cite{Kane13}
\begin{equation}
	\begin{aligned}
		\boldsymbol{R}_{\ssm T}  &= \sum_{i=1}^3 m_i \boldsymbol{T}_i,  \\
		m_i &= \frac{1}{2\pi i}  \int_0^1 d\xi \frac{d}{d\xi} \log\det Q(\xi\boldsymbol{b}_i + \boldsymbol{k}_\perp),
	\end{aligned}
\end{equation}
where $Q(\boldsymbol{k})$ is the equilibrium matrix, $\boldsymbol{b}_i $ are the reciprocal lattice vectors defined through $\boldsymbol{b}_i\cdot \boldsymbol{T}_j = 2\pi \delta_{ij}$ and $\boldsymbol{k}_\perp\cdot \boldsymbol{b}_i=0  $. Distorting the lattice according to Eq.~(\ref{eq:distortion}) ensures that the two oppositely charged Weyl lines along the $(1,1,1)$ direction lie on top of each other \cite{Stenull16}, making $m_i$ independent of $\boldsymbol{k}_\perp$.

The local dipole moment $\boldsymbol{R}_{\ssm L}$ depends on the choice of the unit cell which must be compatible with the surface under consideration. Going through the different configurations we find
\begin{equation}
	\begin{aligned}
		\boldsymbol{R}_{\ssm L}^{\;z,\text{top}} &= s(-1,0,2)\,, & \boldsymbol{R}_{\ssm L}^{\;z,\text{bottom}} &= s(-1,0,-2)\,, \\
		\boldsymbol{R}_{\ssm L}^{\;y,\text{top}} &= s(0,2,-1)\,, & \boldsymbol{R}_{\ssm L}^{\;y,\text{bottom}} &= s(0,-2,-1)\,,
	\end{aligned}
\end{equation}
while $\boldsymbol{R}_{\ssm T}=s(-1,0,2)$. All in all, this results in four and zero (two and two) zero modes for the top and bottom surface in $z$ ($y$) direction. 

For the effective model analysis and manufacturing of the samples, we used an enlarged unit cell with lattice vectors $\tilde{\boldsymbol{T}}_1 = s(2,0,0)$, $\tilde{\boldsymbol{T}}_2 = s(0,2,0)$, $\tilde{\boldsymbol{T}}_3 = s(0,0,2)$, to facilitate being compatible with the different boundaries. This unit cell is four times larger than the original one and hosts two initial unit cells at a given surface, leading to a doubling of the zero mode count as observed in Fig.~\ref{fig:theory}.

%TC:endignore 


\begin{thebibliography}{18}
\expandafter\ifx\csname natexlab\endcsname\relax\def\natexlab#1{#1}\fi
\expandafter\ifx\csname bibnamefont\endcsname\relax
  \def\bibnamefont#1{#1}\fi
\expandafter\ifx\csname bibfnamefont\endcsname\relax
  \def\bibfnamefont#1{#1}\fi
\expandafter\ifx\csname citenamefont\endcsname\relax
  \def\citenamefont#1{#1}\fi
\expandafter\ifx\csname url\endcsname\relax
  \def\url#1{\texttt{#1}}\fi
\expandafter\ifx\csname urlprefix\endcsname\relax\def\urlprefix{URL }\fi
\providecommand{\bibinfo}[2]{#2}
\providecommand{\eprint}[2][]{\url{#2}}

\bibitem[{\citenamefont{Miyamoto et~al.}(1999)\citenamefont{Miyamoto, Kaysser,
  Rabin, Kawasaki, and Ford}}]{FGMBook99}
\bibinfo{editor}{\bibfnamefont{Y.}~\bibnamefont{Miyamoto}},
  \bibinfo{editor}{\bibfnamefont{W.~A.} \bibnamefont{Kaysser}},
  \bibinfo{editor}{\bibfnamefont{B.~H.} \bibnamefont{Rabin}},
  \bibinfo{editor}{\bibfnamefont{A.}~\bibnamefont{Kawasaki}}, \bibnamefont{and}
  \bibinfo{editor}{\bibfnamefont{R.~G.} \bibnamefont{Ford}}, eds.,
  \emph{\bibinfo{title}{Functionally Graded Materials: Design, Processing and
  Applications}} (\bibinfo{publisher}{Springer Science+Business Media},
  \bibinfo{year}{1999}), \href{http://dx.doi.org/}{URL}.

\bibitem[{\citenamefont{Ishikawa et~al.}(2002)\citenamefont{Ishikawa, Yamaoko,
  Harada, Fujii, and Nagasawa}}]{Ishikawa02}
\bibinfo{author}{\bibfnamefont{T.}~\bibnamefont{Ishikawa}},
  \bibinfo{author}{\bibfnamefont{H.}~\bibnamefont{Yamaoko}},
  \bibinfo{author}{\bibfnamefont{Y.}~\bibnamefont{Harada}},
  \bibinfo{author}{\bibfnamefont{T.}~\bibnamefont{Fujii}}, \bibnamefont{and}
  \bibinfo{author}{\bibfnamefont{T.}~\bibnamefont{Nagasawa}}, \emph{A general
  process for in situ formation of functional surface layers on ceramics},
  \bibinfo{journal}{Nature} \textbf{\bibinfo{volume}{416}}, \bibinfo{pages}{64}
  (\bibinfo{year}{2002}), \href{http://dx.doi.org/10.1038/416064a}{URL}.

\bibitem[{\citenamefont{Bartlett et~al.}(2015)\citenamefont{Bartlett, Tolley,
  Overvelde, Weaver, Mosadegh, Bertoldi, Whitesides, and Wood}}]{Bartlett15}
\bibinfo{author}{\bibfnamefont{N.~W.} \bibnamefont{Bartlett}},
  \bibinfo{author}{\bibfnamefont{M.~T.} \bibnamefont{Tolley}},
  \bibinfo{author}{\bibfnamefont{J.~T.~B.} \bibnamefont{Overvelde}},
  \bibinfo{author}{\bibfnamefont{J.~C.} \bibnamefont{Weaver}},
  \bibinfo{author}{\bibfnamefont{B.}~\bibnamefont{Mosadegh}},
  \bibinfo{author}{\bibfnamefont{K.}~\bibnamefont{Bertoldi}},
  \bibinfo{author}{\bibfnamefont{G.~M.} \bibnamefont{Whitesides}},
  \bibnamefont{and} \bibinfo{author}{\bibfnamefont{R.~J.} \bibnamefont{Wood}},
  \emph{A 3D-printed functionally graded soft robot powered by combustion},
  \bibinfo{journal}{Science} \textbf{\bibinfo{volume}{349}},
  \bibinfo{pages}{161} (\bibinfo{year}{2015}),
  \href{http://dx.doi.org/10.1126/science.aab0129}{URL}.

\bibitem[{\citenamefont{Ashby}(2006)}]{ashby2006properties}
\bibinfo{author}{\bibfnamefont{M.}~\bibnamefont{Ashby}}, \emph{The properties
  of foams and lattices}, \bibinfo{journal}{Philosophical Transactions of the
  Royal Society of London A: Mathematical, Physical and Engineering Sciences}
  \textbf{\bibinfo{volume}{364}}, \bibinfo{pages}{15} (\bibinfo{year}{2006}).

\bibitem[{\citenamefont{Schaedler et~al.}(2011)\citenamefont{Schaedler,
  Jacobsen, Torrents, Sorensen, Lian, Greer, Valdevit, and
  Carter}}]{schaedler2011ultralight}
\bibinfo{author}{\bibfnamefont{T.~A.} \bibnamefont{Schaedler}},
  \bibinfo{author}{\bibfnamefont{A.~J.} \bibnamefont{Jacobsen}},
  \bibinfo{author}{\bibfnamefont{A.}~\bibnamefont{Torrents}},
  \bibinfo{author}{\bibfnamefont{A.~E.} \bibnamefont{Sorensen}},
  \bibinfo{author}{\bibfnamefont{J.}~\bibnamefont{Lian}},
  \bibinfo{author}{\bibfnamefont{J.~R.} \bibnamefont{Greer}},
  \bibinfo{author}{\bibfnamefont{L.}~\bibnamefont{Valdevit}}, \bibnamefont{and}
  \bibinfo{author}{\bibfnamefont{W.~B.} \bibnamefont{Carter}}, \emph{Ultralight
  metallic microlattices}, \bibinfo{journal}{Science}
  \textbf{\bibinfo{volume}{334}}, \bibinfo{pages}{962} (\bibinfo{year}{2011}).

\bibitem[{\citenamefont{Zheng et~al.}(2014)\citenamefont{Zheng, Lee,
  Weisgraber, Shusteff, DeOtte, Duoss, Kuntz, Biener, Ge, Jackson
  et~al.}}]{zheng2014ultralight}
\bibinfo{author}{\bibfnamefont{X.}~\bibnamefont{Zheng}},
  \bibinfo{author}{\bibfnamefont{H.}~\bibnamefont{Lee}},
  \bibinfo{author}{\bibfnamefont{T.~H.} \bibnamefont{Weisgraber}},
  \bibinfo{author}{\bibfnamefont{M.}~\bibnamefont{Shusteff}},
  \bibinfo{author}{\bibfnamefont{J.}~\bibnamefont{DeOtte}},
  \bibinfo{author}{\bibfnamefont{E.~B.} \bibnamefont{Duoss}},
  \bibinfo{author}{\bibfnamefont{J.~D.} \bibnamefont{Kuntz}},
  \bibinfo{author}{\bibfnamefont{M.~M.} \bibnamefont{Biener}},
  \bibinfo{author}{\bibfnamefont{Q.}~\bibnamefont{Ge}},
  \bibinfo{author}{\bibfnamefont{J.~A.} \bibnamefont{Jackson}},
  \bibnamefont{et~al.}, \emph{Ultralight, ultrastiff mechanical metamaterials},
  \bibinfo{journal}{Science} \textbf{\bibinfo{volume}{344}},
  \bibinfo{pages}{1373} (\bibinfo{year}{2014}).

\bibitem[{\citenamefont{Meza et~al.}(2014)\citenamefont{Meza, Das, and
  Greer}}]{meza2014strong}
\bibinfo{author}{\bibfnamefont{L.~R.} \bibnamefont{Meza}},
  \bibinfo{author}{\bibfnamefont{S.}~\bibnamefont{Das}}, \bibnamefont{and}
  \bibinfo{author}{\bibfnamefont{J.~R.} \bibnamefont{Greer}}, \emph{Strong,
  lightweight, and recoverable three-dimensional ceramic nanolattices},
  \bibinfo{journal}{Science} \textbf{\bibinfo{volume}{345}},
  \bibinfo{pages}{1322} (\bibinfo{year}{2014}).

\bibitem[{\citenamefont{Kane and Lubensky}(2013)}]{Kane13}
\bibinfo{author}{\bibfnamefont{C.~L.} \bibnamefont{Kane}} \bibnamefont{and}
  \bibinfo{author}{\bibfnamefont{T.~C.} \bibnamefont{Lubensky}},
  \emph{Topological boundary modes in isostatic lattices},
  \bibinfo{journal}{Nature Phys.} \textbf{\bibinfo{volume}{10}},
  \bibinfo{pages}{39} (\bibinfo{year}{2013}),
  \href{http://dx.doi.org/10.1038/nphys2835}{URL}.

\bibitem[{\citenamefont{Huber}(2016)}]{Huber16}
\bibinfo{author}{\bibfnamefont{S.~D.} \bibnamefont{Huber}}, \emph{Topological
  mechanics}, \bibinfo{journal}{Nature Phys.} \textbf{\bibinfo{volume}{12}},
  \bibinfo{pages}{621} (\bibinfo{year}{2016}),
  \href{http://dx.doi.org/10.1038/nphys3801}{URL}.

\bibitem[{\citenamefont{Stenull et~al.}(2016)\citenamefont{Stenull, Kane, and
  Lubensky}}]{Stenull16}
\bibinfo{author}{\bibfnamefont{O.}~\bibnamefont{Stenull}},
  \bibinfo{author}{\bibfnamefont{C.~L.} \bibnamefont{Kane}}, \bibnamefont{and}
  \bibinfo{author}{\bibfnamefont{T.~C.} \bibnamefont{Lubensky}},
  \emph{Topological Phonons and Weyl Lines in Three Dimensions},
  \bibinfo{journal}{Phys. Rev. Lett.} \textbf{\bibinfo{volume}{117}},
  \bibinfo{pages}{068001} (\bibinfo{year}{2016}),
  \href{http://dx.doi.org/10.1103/PhysRevLett.117.068001}{URL}.

\bibitem[{\citenamefont{S{\"u}sstrunk and Huber}(2016)}]{Susstrunk16}
\bibinfo{author}{\bibfnamefont{R.}~\bibnamefont{S{\"u}sstrunk}}
  \bibnamefont{and} \bibinfo{author}{\bibfnamefont{S.~D.} \bibnamefont{Huber}},
  \emph{Classification of topological phonons in linear mechanical
  metamaterials}, \bibinfo{journal}{Proc. Natl. Acad. Sci. USA}
  \textbf{\bibinfo{volume}{113}}, \bibinfo{pages}{E4767}
  (\bibinfo{year}{2016}),
  \href{http://dx.doi.org/10.1073/pnas.1605462113}{URL}.

\bibitem[{\citenamefont{Po et~al.}(2016)\citenamefont{Po, Bahri, and
  Vishwanath}}]{Po16}
\bibinfo{author}{\bibfnamefont{H.~C.} \bibnamefont{Po}},
  \bibinfo{author}{\bibfnamefont{Y.}~\bibnamefont{Bahri}}, \bibnamefont{and}
  \bibinfo{author}{\bibfnamefont{A.}~\bibnamefont{Vishwanath}}, \emph{Phonon
  analog of topological nodal semimetals}, \bibinfo{journal}{Phys. Rev. B}
  \textbf{\bibinfo{volume}{93}}, \bibinfo{pages}{205158}
  (\bibinfo{year}{2016}),
  \href{https://doi.org/10.1103/PhysRevB.93.205158}{URL}.

\bibitem[{\citenamefont{Yang and Zhang}(2016)}]{Yang16}
\bibinfo{author}{\bibfnamefont{Z.}~\bibnamefont{Yang}} \bibnamefont{and}
  \bibinfo{author}{\bibfnamefont{B.}~\bibnamefont{Zhang}}, \emph{Acoustic Weyl
  nodes from stacking dimerized chains}, \bibinfo{journal}{arXiv:1601.07966}
  (\bibinfo{year}{2016}), \href{http://arxiv.org/abs/1601.07966}{URL}.

\bibitem[{\citenamefont{Rocklin et~al.}(2016)\citenamefont{Rocklin, Chen, Falk,
  Vitelli, and Lubensky}}]{Rocklin16}
\bibinfo{author}{\bibfnamefont{D.~Z.} \bibnamefont{Rocklin}},
  \bibinfo{author}{\bibfnamefont{B.~G.} \bibnamefont{Chen}},
  \bibinfo{author}{\bibfnamefont{M.}~\bibnamefont{Falk}},
  \bibinfo{author}{\bibfnamefont{V.}~\bibnamefont{Vitelli}}, \bibnamefont{and}
  \bibinfo{author}{\bibfnamefont{T.~C.} \bibnamefont{Lubensky}},
  \emph{Mechanical Weyl Modes in Topological Maxwell Lattices},
  \bibinfo{journal}{Phys. Rev. Lett.} \textbf{\bibinfo{volume}{116}},
  \bibinfo{pages}{135503} (\bibinfo{year}{2016}),
  \href{https://doi.org/10.1103/PhysRevLett.116.135503}{URL}.

\bibitem[{\citenamefont{Bernevig}(2015)}]{Bernevig15}
\bibinfo{author}{\bibfnamefont{B.~A.} \bibnamefont{Bernevig}}, \emph{It's been
  a Weyl coming}, \bibinfo{journal}{Nature Phys.}
  \textbf{\bibinfo{volume}{11}}, \bibinfo{pages}{698} (\bibinfo{year}{2015}),
  \href{http://dx.doi.org/10.1038/nphys3454}{URL}.

\bibitem[{\citenamefont{Maxwell}(1864)}]{Maxwell64}
\bibinfo{author}{\bibfnamefont{J.~C.} \bibnamefont{Maxwell}}, \emph{On the
  calculation of the equilibrium and stiffness of frames},
  \bibinfo{journal}{Phil. Mag.} \textbf{\bibinfo{volume}{27}},
  \bibinfo{pages}{294} (\bibinfo{year}{1864}).

\bibitem[{\citenamefont{Calladine}(1978)}]{Calladine78}
\bibinfo{author}{\bibfnamefont{C.~R.} \bibnamefont{Calladine}},
  \emph{Buckminster Fuller's ``Tensegrity'' structures and Clerk Maxwell's
  rules for the construction of stiff frames}, \bibinfo{journal}{Int. J. Solids
  Struct.} \textbf{\bibinfo{volume}{14}}, \bibinfo{pages}{161}
  (\bibinfo{year}{1978}),
  \href{http://dx.doi.org/10.1016/0020-7683(78)90052-5}{URL}.

\bibitem[{\citenamefont{Lubensky et~al.}(2015)\citenamefont{Lubensky, Kane,
  Mao, Souslov, and Sun}}]{Lubensky15}
\bibinfo{author}{\bibfnamefont{T.~C.} \bibnamefont{Lubensky}},
  \bibinfo{author}{\bibfnamefont{C.~L.} \bibnamefont{Kane}},
  \bibinfo{author}{\bibfnamefont{X.}~\bibnamefont{Mao}},
  \bibinfo{author}{\bibfnamefont{A.}~\bibnamefont{Souslov}}, \bibnamefont{and}
  \bibinfo{author}{\bibfnamefont{K.}~\bibnamefont{Sun}}, \emph{Phonons and
  elasticity in critically coordinated lattices}, \bibinfo{journal}{Rep. Prog.
  Phys.} \textbf{\bibinfo{volume}{78}}, \bibinfo{pages}{109501}
  (\bibinfo{year}{2015}),
  \href{http://dx.doi.org/10.1088/0034-4885/78/7/073901}{URL}.

\end{thebibliography}
\end{document}